\documentclass{ieeeaccess}
\usepackage{cite}
\usepackage{amsmath,amssymb,amsfonts}
\usepackage{algorithmic}
\usepackage{bm}
\usepackage{graphicx}
\usepackage{braket}
\usepackage{comment}
\usepackage{textcomp}

\usepackage{subfigure}
\usepackage{booktabs}
\usepackage{framed}

\def\BibTeX{{\rm B\kern-.05em{\sc i\kern-.025em b}\kern-.08em
    T\kern-.1667em\lower.7ex\hbox{E}\kern-.125emX}}
\begin{document}
\history{Date of publication xxxx 00, 0000, date of current version xxxx 00, 0000.}
\doi{10.1109/ACCESS.2017.DOI}

% For submit to arXiv
\onecolumn
\begin{framed}
   \noindent
   This work has been submitted to the IEEE for possible publication. Copyright may be transferred without notice, after which this version may no longer be accessible.%
\end{framed}
\clearpage
\twocolumn

\title{Inductive Construction of Variational Quantum Circuit for Constrained Combinatorial Optimization}
\author{\uppercase{Hyakka Nakada}\authorrefmark{1,2},
\uppercase{Kotaro Tanahashi\authorrefmark{1}, 
and Shu Tanaka}\authorrefmark{2,3,4,5}, \IEEEmembership{Member, IEEE}}

\address[1]{Recruit Co., Ltd., Tokyo 100-6640, Japan}
\address[2]{Graduate School of Science and Technology, Keio University, Kanagawa 223-8522, Japan}
\address[3]{Department of Applied Physics and Physico-Informatics, Keio University, Kanagawa 223-8522, Japan}
\address[4]{Keio University Sustainable Quantum Artificial Intelligence Center (KSQAIC), Keio University, Tokyo 108-8345, Japan}
\address[5]{Human Biology-Microbiome-Quantum Research Center (WPI-Bio2Q), Keio University, Tokyo 108-8345, Japan}

\tfootnote{This work was partially supported by JSPS KAKENHI (Grant Number JP23H05447), the Council for Science, Technology, and Innovation (CSTI) through the Cross-ministerial Strategic Innovation Promotion Program (SIP), ``Promoting the application of advanced quantum technology platforms to social issues'' (Funding agency: QST), JST (Grant Number JPMJPF2221).} 

\markboth
{H. Nakada \headeretal: Inductive Construction of Variational Quantum Circuit for Constrained Combinatorial Optimization}
{H. Nakada \headeretal: Inductive Construction of Variational Quantum Circuit for Constrained Combinatorial Optimization}

\corresp{Corresponding author: Hyakka Nakada (e-mail: hyakka\_nakada@r.recruit.co.jp).}

\begin{abstract}
In this study, we propose a new method for constrained combinatorial optimization using variational quantum circuits. Quantum computers are considered to have the potential to solve large combinatorial optimization problems faster than classical computers. Variational quantum algorithms, such as Variational Quantum Eigensolver (VQE), have been studied extensively because they are expected to work on noisy intermediate scale devices. Unfortunately, many optimization problems have constraints, which induces infeasible solutions during VQE process. Recently, several methods for efficiently solving constrained combinatorial optimization problems have been proposed by designing a quantum circuit so as to output only the states that satisfy the constraints. However, the types of available constraints are still limited. Therefore, we have started to develop variational quantum circuits that can handle a wider range of constraints. 
The proposed method utilizes a forwarding operation that maps from feasible states for subproblems to those for larger subproblems. As long as appropriate forwarding operations can be defined, iteration of this process can inductively construct variational circuits outputting feasible states even in the case of multiple and complex constraints. 
In this paper, the proposed method was applied to facility location problem and was found to increase the probability for measuring feasible solutions or optimal solutions. In addition, the cost of the obtained circuit was comparable to that of conventional variational circuits. 

\end{abstract}


\titlepgskip=-15pt

\maketitle

\section{Introduction}
\label{sec:introduction}
\Figure[t!](topskip=0pt, botskip=0pt, midskip=0pt)[width=16.5cm]{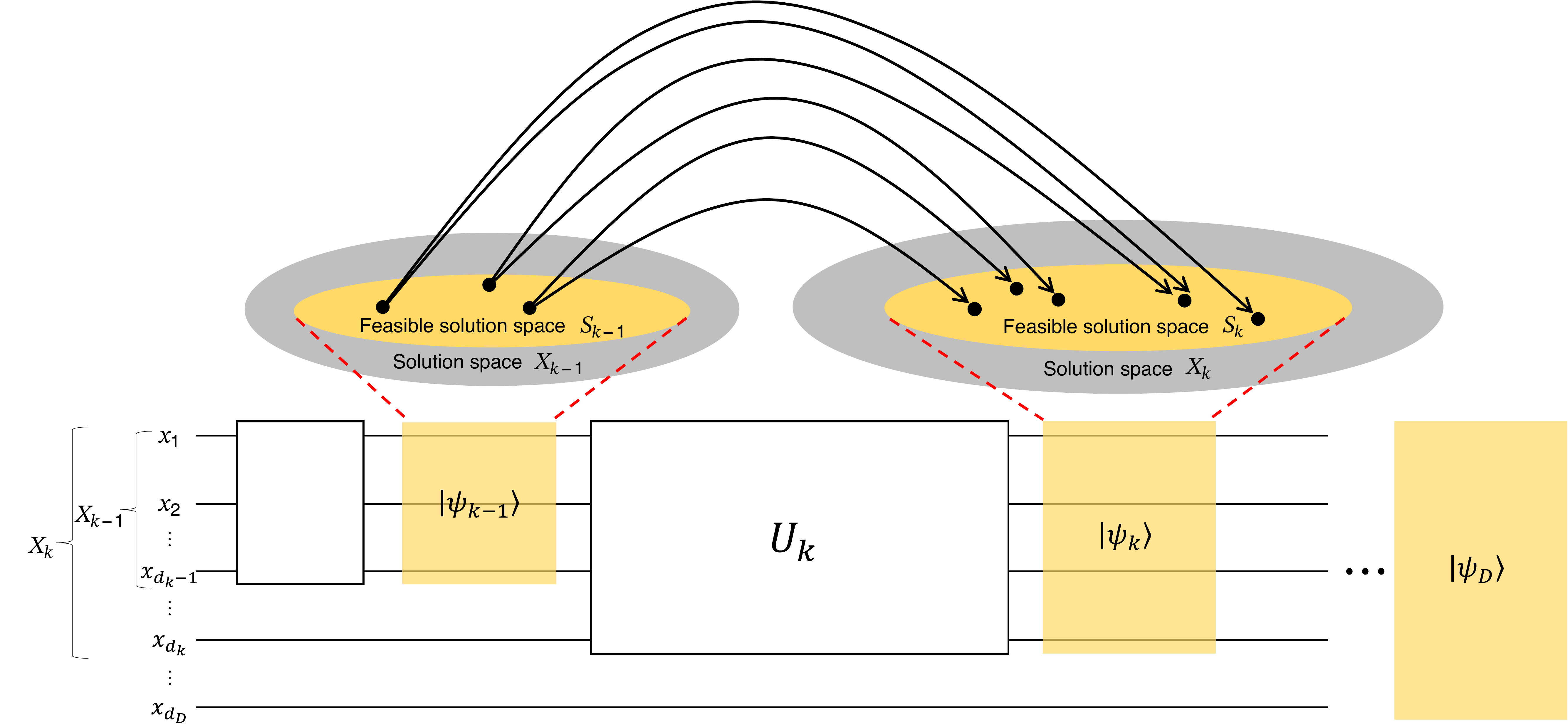}
{Schematic picture of proposed method. The intermediate anstaz $\left| \psi_{k-1}\right\rangle$ outputs the states superposintioned by feasible solutions of $(k-1)$th subproblem. Forwarding operation $U_k$ is designed to map from the feasible states for the subproblems to that for the larger subproblems. 
Repeating this operation can construct variational circuits outputting feasible states in original problems.
\label{fig:fig1}}
\PARstart{C}{ombinatorial} optimization problems have many real-world applications. In recent years, the size of these problems has increased with the volume of data traffic, which leads to the difficulty of solving them in a realistic time. 
To address this situation, not only heuristics for approximate solutions but also quantum computing technology have been actively developed. 

Unlike classical computers, quantum computers have the property of simultaneously possessing the states of $0$ and $1$. Adding appropriate quantum operations to such a superpositioned state can increase the amplitude of the states for good solutions.
Thus, quantum algorithms are expected to solve large combinatorial optimization problems at high speed.
Quantum Annealing (QA)~\cite{kadowaki1998ae, farhi2000ae} is known as an optimization method to find the global minimum of an objective function, by a process using quantum fluctuations. 

In addition, there are several quantum-gate based methods.
For example, optimization using Grover adaptive search~\cite{gilliam2019gas} or quantum phase estimation~\cite{lee2023qpe} has been proposed. 
These methods essentially require fault-tolerant quantum computing.
Quantum-classical hybrid algorithms using Variational Quantum Circuit (VQC)~\cite{moll2018vqa} have been proposed. This method is expected to work with current Noisy Intermediate-Scale Quantum devices (NISQ)~\cite{preskil2018nisq} and has been actively investigated. This method parameterizes a quantum circuit with variational parameters. Based on the expected energy of cost Hamiltonian corresponding to an objective function, the variational parameters are optimized so as to minimize this value. When the variational parameters are well learned, the variational circuit outputs the states for good solutions. Variational Quantum Eigensolver (VQE)~\cite{peruzzo2014vqe, mcclean2016vqe, wang2019vqe, parrish2019vqe, kandala2017vqe} and Quantum Approximate Optimization Algorithm (QAOA)~\cite{farhi2014quantum, choi2019quantum} are typical examples of such methods.

Many combinatorial optimization problems are constrained so that optimal solutions are strictly selected from feasible solution space. In conventional methods using quantum gates, feasible solutions are preferentially searched for by minimizing the customized objective function with a penalty term for the constraints~\cite{koretsky2021penalty, brandhofer2022penalty}. That is, penalty function methods are usually utilized~\cite{bertsekas1982penalty, luenberger2008penalty, lucas2014penalty, tanaka2017penalty, takehara2019penalty, tamura2021penalty, tanahashi2019penalty,
zaman2022penalty, gao2023quantum}. However, these methods raise several challenges. An increase in the number of interactions in a Hamiltonian may disable performing computation in NISQ devices. In addition, the penalty coefficients are not straightforward to adjust appropriately, and emergence of infeasible solutions is principally inevitable. 
Therefore, without penalty function methods, noble approaches have been developed to search for feasible solutions by designing specific VQCs. For example, several studies on QAOA have reported that mixer gates are modified so that transitions occur only between the states of feasible solutions~\cite{hadfield2019qaoa, wang2020qaoa, bartschi2020grover, christiansen2024qaoa}. 
In addition, whereas conventional VQE utilizes simple VQCs multilayered by rotating gates and controlled-NOT (CNOT) gates~\cite{amaro2022jss}, several studies tried to establish specific VQCs so that their output states are exactly feasible~\cite{matsuo2023pqc, quinones2020pqc}. By restricting the space of output states to that of the feasible solution space, the region to be searched can be reduced exponentially, which enables efficient optimization. 
Hereinafter, we focus on VQE algorithms without penalty function methods.

Previous studies have derived VQCs that output only feasible states for one-hot constraints~\cite{matsuo2023pqc}, comparison inequality constraints~\cite{quinones2020pqc}, and constraints specialized for traveling salesman problems~\cite{matsuo2023pqc}. 
However, few efforts have been made to systematically design appropriate VQCs for more general types of constraints in the field of VQE. This may be due to the following reasons. It is not trivial to construct ansatzes that handle even a single constraint. Although, in principle, Grover's algorithm can prepare the states that satisfy the constraints by introducing desirable oracles, the circuit becomes too complex and expensive to be employed as a VQC. Therefore, a hardware-efficient ansatz must be derived from scratch and through trial and error. In addition, constructing ansatzes is much more difficult in the case of multiple constraints. 
Supposing that appropriate ansatzes are obtained for each single constraint, VQCs for all constraints cannot be constructed by combining their circuits in series.
This is because the ansatz for each constraint is originally designed to act on the initialized state $\left| 0\right\rangle$.

To construct VQCs systematically, we propose a method to generate a quantum circuit inductively by adding a sequential gate operation to the intermediate ansatz for a subproblem.
This operation is hereinafter referred to as forwarding operation, which transfers the states of feasible solutions for a present subproblem into those for a larger problem, as shown in Fig.~\ref{fig:fig1}.
By repeating the forwarding operations a given number of times, the obtained circuit can output the states of feasible solutions for the original problem. This inductive method can be applied to any problems for which the forwarding operation ensures the ergodicity in the feasible solution space.
Thus, it has the potential to construct VQCs for a broader type of constraint. 
In a related research~\cite{matsuo2023pqc}, VQCs specialized for traveling salesman problems have been derived by utilizing the property of permutation matrices. Our method generalizes this procedure. 

The remainder of this paper is organized as follows. In Section~\ref{sec:preliminaries}, an overview of the related method~\cite{matsuo2023pqc} is given. In Section~\ref{sec:inductive}, a generalization of the method is explained, followed by examples of VQCs for several specific problems. 
We evaluated the cost of the quantum circuits constructed by the proposed method and compared it with the conventional methods in Section~\ref{sec:cost}. In Section~\ref{sec:exp}, we performed numerical calculations for the facility location problem. Finally, Section~\ref{sec:summary} summarizes the paper.

\section{Preliminaries}
\label{sec:preliminaries}

Subsection~\ref{subsec:feasibleQC} explains the concept of a feasible quantum circuit, a pivotal concept in this paper. 
In Subsection~\ref{subsec:related}, we introduce previous studies related to feasible VQC for constrained combinatorial optimization to clarify the contribution of this study.

\subsection{Feasible Quantum Circuit}
\label{subsec:feasibleQC}
We consider a quantum circuit that outputs only states superposintioned by at least one feasible solution for given constraints and does not output any states that violate the constraints. This circuit is defined as a ``feasible'' quantum circuit for the constraints. 
Additionally, a special feasible quantum circuit that can output states superposintioned by all feasible solutions is called ``fully feasible'' quantum circuit. This paper focuses on constructing fully feasible VQCs. 

\subsection{Related Study on Feasible VQC for Constrained Combinatorial Optimization}
\label{subsec:related}
\Figure[t!](topskip=0pt, botskip=0pt, midskip=0pt)[width=8.5cm]{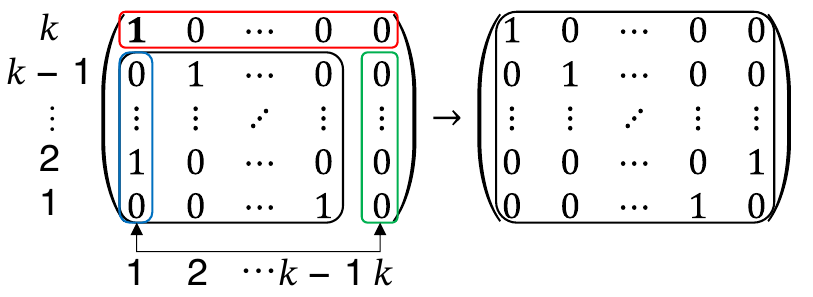}
{Generating fully feasible VQC for the traveling salesman problem~\cite{matsuo2023pqc}. A $(k-1) \times (k-1)$ permutation matrix is transferred to a $k \times k$ permutation matrix. The corresponding quantum-gate operations are performed sequentially.
\label{fig:fig2}}

As the related study, a fully feasible VQC has been derived for the traveling salesman problem~\cite{matsuo2023pqc}.
Let $n$ be the total number of city vertices.
$x_{i,j}\ \text{for}\ i,j=1,2,\ldots,n$ are binary variables where $i$ represents the city vertex and $j$ represents its order in a prospective cycle.
If the $i$th city is visited at the $j$th step, $x_{i,j}=1$, otherwise $x_{i,j}=0$.
As the constraints, a group of conditions 
\begin{equation}
\label{eq:eq1}
\sum_{i=1}^{n}x_{i,j} =1 \ \ \text{for} \ j=1,2,\ldots,n,
\end{equation}
\begin{equation}
\label{eq:eq2}
\sum_{j=1}^{n}x_{i,j} =1 \ \ \text{for} \ i=1,2,\ldots,n
\end{equation}
is imposed. 
Let $k$ be a natural number less than or equal to $n$.
We consider the case where $n\to k$ in the constraints~\eqref{eq:eq1} and \eqref{eq:eq2}.  
Such a subproblem is to find solutions satisfying $\sum_{i=1}^{k} x_{i,j} =\sum_{j=1}^{k} x_{i,j} =1 \ \text{for}\ i,j=1,2,\ldots,k$ for the variables $x_{1,1},\ldots,x_{k,k}$.
Let $\left| \psi_k\right\rangle$ be a fully feasible ansatz for this subproblem and $q_{1,1},\ldots,q_{k,k}$ be the qubits corresponding to the variables $x_{1,1},\ldots,x_{k,k}$.
These qubits are rearranged by the following matrix form 
\begin{equation}
\label{eq:eq3}
\left(\begin{matrix}
q_{1,k}&q_{2,k}&\cdots&q_{k-1,k}&q_{k,k}\\
q_{1,k-1}&q_{2,k-1}&\cdots&q_{k-1,k-1}&q_{k,k-1}\\
\vdots&\vdots&\ddots&\vdots&\vdots\\
q_{1,2}&q_{2,2}&\cdots&q_{k-1,2}&q_{k,2}\\
q_{1,1}&q_{2,1}&\cdots&q_{k-1,1}&q_{k,1}
\end{matrix}\right).
\end{equation}
Then, the constraints are equivalent to the condition that the matrix~\eqref{eq:eq3} is a permutation matrix.
This is because the permutation matrix is a square binary matrix that has exactly one entry of $1$ in each row and each column with all other entries $0$.
There is a recursive relation between the $(k-1)\times (k-1)$ and $k\times k$ permutation matrices. In the left of Fig.~\ref{fig:fig2}, one of the $(k-1)\times (k-1)$ permutation matrices is given inside the black frame. An one-hot bitstring of length $k$ is added, as shown inside the red frame. Then, an all-zeros bitstring of length $k-1$ as shown inside the green frame. 
In Fig.~\ref{fig:fig2}, the specific case where $q_{1,k}=1$ is depicted.
The constraint $\sum_{j=1}^k x_{1,j} =1$ is violated because there are two elements with the value of $1$ in the first column. By swapping the bitstring inside the blue frame and that inside the green frame, this constraint is recovered with the other constraints preserved.
Thus, the $k\times k$ permutation matrix can be obtained as shown in the right of Fig.~\ref{fig:fig2}. Implementing the above operation with quantum gates, the fully feasible ansatz $\left| \psi_k\right\rangle$ for the larger problem can be generated from $\left| \psi_{k-1}\right\rangle$. 

The explicit gates generating $\left| \psi_{k}\right\rangle$ from $\left| \psi_{k-1}\right\rangle$ are explained below. 
First, a one-hot state is prepared on $q_{1,k},q_{2,k},\ldots,q_{k,k}$. As the equally weighted one-hot state, a W state is well known~\cite{diker2022wstate}. Instead, to construct a VQC, the parameterized W state whose coefficients are parameterized by variational parameters has been adopted~\cite{matsuo2023pqc}. For the details of the parameterized W state, refer to Appendix~\ref{sec:pws}. 
Then, $q_{k,1},q_{k,2},\ldots,q_{k,k}$ are initialized to $\left| 0\right\rangle$.
After $q_{u,k} \ \text{for}\ u=1,2,\ldots,k-1$ are set as control bits, controlled-SWAP (CSWAP) gates~\cite{fredkin1982cswap} act on $q_{u,v}$ and $q_{k,v}$ for $v=1,2,\ldots,k-1$. These gates can generate $\left| \psi_{k}\right\rangle$ from $\left| \psi_{k-1}\right\rangle$.
Because $\left| \psi_1\right\rangle=\left| q_{1,1}=1\right\rangle$ can be obtained by applying an $X$ gate to $q_{1,1}$, performing the above operations up to $k=n$ creates the target state $\left| \psi_ n\right\rangle$.

\section{Inductive Construction of Feasible VQC}
\label{sec:inductive}
In Subsection~\ref{subsec:generalization}, we generalize the related method in Subsection~\ref{subsec:related}.
From Subsection~\ref{subsec:feasibleVQC_assignment}, fully feasible VQCs are constructed by proposed method for several typical problems: assignment problem, shift scheduling problem, and facility location problem. Our study is the first study to derive fully feasible VQCs for those problems.

\subsection{Generalization of Related Study}
\label{subsec:generalization}
We propose a method to generate a VQC inductively by adding a forwarding operation to the intermediate ansatz for a subproblem, as shown in Fig.~\ref{fig:fig1}.
Let $\left| \psi_k\right\rangle$ be a fully feasible ansatz for a $k$-dimensional subproblem, and $D$ be the dimension of the original problem.
Hereinafter, such a subproblem is denoted by the $k$th subproblem.
This ansatz may have auxiliary qubits $a_1,\ldots,a_{e_k}$ as well as the original qubits $q_1,\ldots,q_{d_k}$ corresponding to the variables of the subproblem $(x_1,\ldots,x_{d_k})\in X_k$. 
Here, $X_k$ represents the solution space for the $k$th subproblem.
In Fig.~\ref{fig:fig1}, the auxiliary qubits are omitted for simplicity. 

Let $S_k$ be the set of feasible solutions for the $k$th subproblem and $\bm{\theta}_k$ be the vector of variational parameters in $\left| \psi_k\right\rangle$. The intermediate ansatz is generally represented by
\begin{equation}
\label{eq:eq4}
\left| \psi_k\right\rangle = \sum_{\bm{q}_k \in S_k} c_{k\bm{q}_k}\left(\bm{\theta}_k \right) \left| q_1,\ldots,q_{d_k} \right\rangle \otimes \left| a_1,\ldots,a_{e_k} \right\rangle,
\end{equation}
where $\bm{q}=(q_1,\ldots,q_{d_k})$. The coefficients $c_{k\bm{q}_k}\left(\bm{\theta}_k \right)$ are parameterized by variational parameters $\bm{\theta}_k$.

We assume that the initial state $\left| \psi_1\right\rangle$ is prepared a priori.
Then, a forwarding operation $U_k$ is applied sequentially to the intermediate ansatz $\left| \psi_{k-1}\right\rangle$.
$\left| \psi_D\right\rangle$ is obtained as the fully feasible ansatz for the original problem.
By measuring the expected energy value, the variational parameters are optimized so that the value is minimized.

In the forwarding operation $U_k$, $q_{d_{k-1}+1},q_{d_{k-1}+2},\ldots,q_{d_{k}}$ and $a_{e_{k-1}+1},a_{e_{k-1}+2},\ldots,a_{e_{k}}$ qubits are added to the quantum register of the ansatz $\left| \psi_{k-1}\right\rangle$.
Next, the existing state of $q_1,\ldots,q_{d_{k-1}}$ transitions to the feasible state in $S_{k}$.
To ensure the full feasibleness, in other words, ergodicity for feasible solution space, the obtained ansatz $\left| \psi_{k}\right\rangle$ must represent all states in $S_{k}$. In the following sections, fully feasible VQCs for several typical problems are constructed by the proposed method.

\subsection{Fully Feasible VQC for Assignment-type Constraint}
\label{subsec:feasibleVQC_assignment}
\Figure[t!](topskip=0pt, botskip=0pt, midskip=0pt)[width=8.5cm]{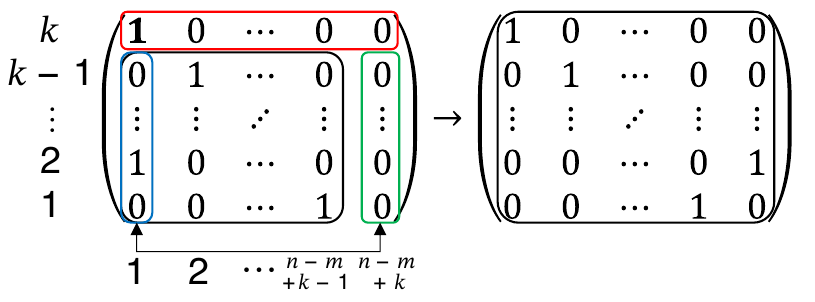}
{Generating fully feasible VQC for assignment problem. A forwarding operation to the $k$th subproblem is depicted. This subproblem is defined by reducing $m\to k, n\to n-m+k$ in the constraints~\eqref{eq:eq5} and \eqref{eq:eq6}.
\label{fig:fig3}}

In the assignment problem, $x_{i,j}$ is a binary variable that takes $1$ if the $i$th worker is assigned to the $j$th job and $0$ otherwise.
Let $m$ and $n$ be the total number of jobs and that of the workers, respectively, where $m \leq n$.
As the typical constraints, a group of conditions 
\begin{equation}
\label{eq:eq5}
\sum_{i=1}^n x_{i,j} =1 \ \ \text{for} \ j=1,2,\ldots,m,
\end{equation}
\begin{equation}
\label{eq:eq6}
\sum_{j=1}^m x_{i,j} \leq 1 \ \ \text{for} \ i=1,2,\ldots,n
\end{equation}
is imposed.
As detailed in Subsection~\ref{subsec:generalization}, we let $\bm{q}$ be the qubits corresponding to the variables $\bm{x}$.

A construction of VQCs for the assignment problem is as follows.
Let $k$ be a natural number less than or equal to $m$.
We consider the case where $m\to k, n\to n-m+k$ in the constraints~\eqref{eq:eq5} and \eqref{eq:eq6}.
A feasible intermediate ansatz for such the $k$th subproblem is represented by $\left| \psi_k\right\rangle$.
Thus, any states outputted from $\left| \psi_k\right\rangle$ must satisfy ($k-1$)th subproblem's constraints $\sum_{i=1}^{n-m+k} q_{i,j} =1 \ \text{for} \ j=1,2,\ldots,k$ and $\sum_{j=1}^k q_{i,j} \leq 1 \ \text{for} \ i=1,2,\ldots,n-m+k$.
In the case of $k=1$, only the first constraint must be satisfied because the second is trivial. 
Thus, $\left| \psi_1\right\rangle$ can be prepared by generating parameterized W state on $q_{1,1},q_{2,1},\ldots,q_{n-m+1,1}$.
Next, the forwarding operation that generates $\left| \psi_k\right\rangle$ from $\left| \psi_{k-1}\right\rangle$ is explained. 
The qubits are rearranged by the following matrix form 
\begin{equation}
\label{eq:eq7}
\left(\begin{matrix}
q_{1,k}&q_{2,k}&\cdots&q_{n-m+k-1,k}&q_{n-m+k,k}\\
q_{1,k-1}&q_{2,k-1}&\cdots&q_{n-m+k-1,k-1}&q_{n-m+k,k-1}\\
\vdots&\vdots&\ddots&\vdots&\vdots\\
q_{1,2}&q_{2,2}&\cdots&q_{n-m+k-1,2}&q_{n-m+k,2}\\
q_{1,1}&q_{2,1}&\cdots&q_{n-m+k-1,1}&q_{n-m+k,1}
\end{matrix}\right).
\end{equation}
In the left of Fig.~\ref{fig:fig3}, $q_{1,1},\ldots,q_{n-m+k-1,k-1}$ satisfying ($k-1$)th subproblem's constraints are given inside the black frame.
An one-hot bitstring ($q_{1,k},q_{2,k},\ldots,q_{n-m+k,k}$) and all-zeros bitstring ($q_{n-m+k,1},\ldots,q_{n-m+k,k-1}$) are added as shown in the red and green frame, respectively.
In Fig.~\ref{fig:fig3}, the specific case where $q_{1,k} = 1$ is depicted.
The second constraint is violated because there are two elements with the value of $1$ in the first column. 
By swapping the bitstring inside the blue frame and that inside the green frame, this constraint is recovered with the other constraints preserved.
Thus, $q_{1,1},\ldots,q_{n-m+k,k}$ can satisfy $k$th subproblem's constraints.

The explicit gates are as follows. 
First, $\left| \psi_1\right\rangle$ is obtained by preparing a parameterized W state on $q_{1,1},q_{2,1},\ldots,q_{n-m+1,1}$. 
Then, forwarding operations are implemented. A parameterized W state is prepared on $q_{1,k},q_{2,k},\ldots,q_{n-m+k,k}$, and $q_{n-m+k,1},\ldots,q_{n-m+k,k-1}$ are initialized to $\left| 0\right\rangle$.
After $q_{u,k} \ \text{for}\ u=1,2,\ldots,n-m+k-1$ are set as control bits, CSWAP gates act on $q_{u,v}$ and $q_{n-m+k,v}$ for $v=1,2,\ldots,k-1$. These gates can generate $\left| \psi_k\right\rangle$.
Performing this forwarding operation up to $k=m$ creates the target state $\left| \psi_ m\right\rangle$, which represents a feasible ansatz for the original problem.

We prove that $\left| \psi_ m\right\rangle$ is fully feasible as follows. 
Let $C_k$ be the number of distinct feasible solutions encoded in $\left| \psi_k\right\rangle$. The $k$th forwarding operation $U_k$ adds a parameterized W state, which is a superposition of $n-m+k$ one-hot states, to each existing state in $\left| \psi_{k-1}\right\rangle$. 
Thus, the number of distinct feasible solutions encoded in $\left| \psi_k\right\rangle$ is $C_k=(n-m+k)C_{k-1}$. Applying this recurrence formula to $C_1=n-m+1$ derives $C_m={}_n P_{m}$.
On the other hand, the number of distinct feasible solutions of the original assignment problem is ${}_n P_{m}$.
Thus, the ansatz $\left| \psi_ m\right\rangle$ can generate states superpositioned by all feasible solutions.

\subsection{Fully Feasible VQC for Shift-scheduling type Constraint}
\label{subsec:feasibleVQC_shift}
\Figure[t!](topskip=0pt, botskip=0pt, midskip=0pt)[width=8.5cm]{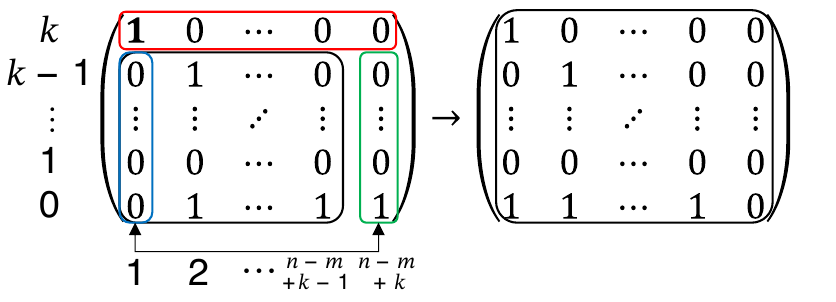}
{Generating fully feasible VQC for shift scheduling problem.
A forwarding operation to the $k$th subproblem is depicted. This subproblem is defined by reducing $m\to k, n\to n-m+k$ in the constraints~\eqref{eq:eq8} and \eqref{eq:eq9}.
The bottom row is the qubits $\bm{r}$ corresponding to the variables $\bm{y}$, and the others are the qubits $\bm{q}$ corresponding to the variables $\bm{x}$.
\label{fig:fig4}}

In the shift scheduling problem, $x_{i,j}$ is a binary variable that takes $1$ if the $i$th worker is assigned to the $j$th shift and $0$ otherwise. 
$y_i$ is a variable that takes $1$ if the $i$th worker is employed and $0$ otherwise. 
Let $m$ and $n$ be the total number of shifts and that of workers, respectively.
As the typical constraints, a group of conditions 
\begin{equation}
\label{eq:eq8}
\sum_{i=1}^n x_{i,j} =1 \ \ \text{for} \ j=1,2,\ldots,m,
\end{equation}
\begin{equation}
\label{eq:eq9}
\sum_{j=1}^m x_{i,j} \leq y_i \ \ \text{for} \ i=1,2,\ldots,n
\end{equation}
is imposed ($m\leq n$).
We let $\bm{q}$ and $\bm{r}$ be the qubits corresponding to the variables $\bm{x}$ and those to the variables $\bm{y}$, respectively.

A construction of VQCs for the shift scheduling problem is as follows.
We consider the case where $m\to k, n\to n-m+k$ in the constraints~\eqref{eq:eq8} and \eqref{eq:eq9}.
$k$ denotes a natural number less than or equal to $m$.
A feasible intermediate ansatz for such the $k$th subproblem is represented by $\left| \psi_k\right\rangle$.
Thus, any states outputted from $\left| \psi_k\right\rangle$ must satisfy ($k-1$)th subproblem's constraints $\sum_{i=1}^{n-m+k} q_{i,j} =1 \ \ \text{for} \ j=1,2,\ldots,k$ and $\sum_{j=1}^k q_{i,j} \leq r_i \ \ \text{for} \ i=1,2,\ldots,n-m+k$.
In the interpolative case of $k=0$, that is the case where the variables $\bm{x}$ are absent, the constraints are trivial. The resolution of constraints ensures that $y_1,y_2,\ldots,y_{n-m}$ have arbitrary values.
Thus, $\left| \psi_0\right\rangle$ can be prepared by acting $R_y (\phi_i)$ gates on $r_{1},r_{2},\ldots,r_{n-m}$. Here, $\phi_i$ is a variational parameter.
Next, the forwarding operation that generates $\left| \psi_k\right\rangle$ from $\left| \psi_{k-1}\right\rangle$ is explained. 
The qubits are rearranged by the following matrix form 
\begin{equation}
\label{eq:eq10}
\left(\begin{matrix}
q_{1,k}&q_{2,k}&\cdots&q_{n-m+k-1,k}&q_{n-m+k,k}\\
q_{1,k-1}&q_{2,k-1}&\cdots&q_{n-m+k-1,k-1}&q_{n-m+k,k-1}\\
\vdots&\vdots&\ddots&\vdots&\vdots\\
q_{1,2}&q_{2,2}&\cdots&q_{n-m+k-1,2}&q_{n-m+k,2}\\
q_{1,1}&q_{2,1}&\cdots&q_{n-m+k-1,1}&q_{n-m+k,1}\\
r_{1}&r_{2}&\cdots&r_{n-m+k-1}&r_{n-m+k}
\end{matrix}\right).
\end{equation}
In the left of Fig.~\ref{fig:fig4}, $q_{1,1},\ldots,q_{n-m+k-1,k-1}$ and $r_1,\ldots,r_{n-m+k-1}$ inside the black frame satisfy ($k-1$)th subproblem's constraints.
An one-hot bitstring ($q_{1,k},q_{2,k},\ldots,q_{n-m+k,k}$) is added as shown in the red frame.
An all-zeros bitstring ($q_{n-m+k,1},\ldots,q_{n-m+k,k-1}$) and all-ones bitstring ($r_{n-m+k}$) are added in the green frame.
In Fig.~\ref{fig:fig4}, the specific case where $q_{1,k} = 1$ is depicted.
The second constraint is violated because $r_1$ remains $\left| 0\right\rangle$.
There are two elements with the value of $1$ in the first column. 
By swapping the bitstring inside the blue frame and that inside the green frame, this constraint is recovered with the other constraints preserved.
Thus, $q_{1,1},\ldots,q_{n-m+k,k},r_1,\ldots,r_{n-m+k}$ can satisfy $k$th subproblem's constraints.

The explicit gates are as follows. 
First, $\left| \psi_0\right\rangle$ is obtained  by acting $R_y (\phi_i)$ gates on $r_{1},r_{2},\ldots,r_{n-m}$. 
Then, forwarding operations are implemented. A parameterized W state is prepared on $q_{1,k},q_{2,k},\ldots,q_{n-m+k,k}$, and $q_{n-m+k,1},\ldots,q_{n-m+k,k-1}$ are initialized to $\left| 0\right\rangle$.
In addition, $r_{n-m+k}$ is initialized to $\left| 1\right\rangle$ with an $X$ gate.
After $q_{u,k} \ \text{for}\ u=1,2,\ldots,n-m+k-1$ are set as control bits, CSWAP gates act on $q_{u,v}$ and $q_{n-m+k,v}$ for $v=1,2,\ldots,k-1$. Similarly, CSWAP gates act on $r_u$ and $r_{n-m+k}$.
These gates can generate $\left| \psi_k\right\rangle$.
Performing this forwarding operation up to $k=m$ creates the target state $\left| \psi_ m\right\rangle$, which represents a feasible ansatz for the original problem.

We prove that $\left| \psi_ m\right\rangle$ is fully feasible as follows. 
Let $C_k$ be the number of distinct feasible solutions encoded in $\left| \psi_k\right\rangle$. The $k$th forwarding operation $U_k$ adds a parameterized W state, which is a superposition of $n-m+k$ one-hot states, to each existing state in $\left| \psi_{k-1}\right\rangle$. 
Thus, the number of distinct feasible solutions encoded in $\left| \psi_k\right\rangle$ is $C_k=(n-m+k)C_{k-1}$. Applying this recurrence formula to $C_0=2^{n-m}$ derives $C_m={}_n P_{m} 2^{n-m}$.
On the other hand, the number of distinct feasible solutions of the original shift scheduling problem is ${}_n P_{m} 2^{n-m}$.
Thus, it is shown that the ansatz $\left| \psi_ m\right\rangle$ can generate states superpositioned by all feasible solutions.

\subsection{Fully Feasible VQC for Facility-location type Constraint}
\label{subsec:feasibleVQC_facility}
\Figure[t!](topskip=0pt, botskip=0pt, midskip=0pt)[width=8.5cm]{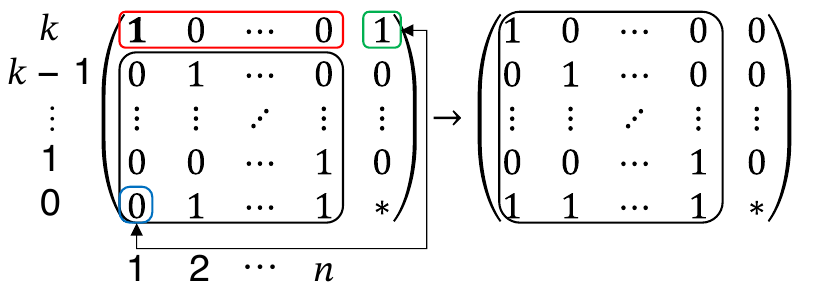}
{Generating fully feasible VQC for facility location problem. 
A forwarding operation to the $k$th subproblem is depicted. This subproblem is defined by reducing $m\to k$ in the constraints~\eqref{eq:eq11} and \eqref{eq:eq12}.
The bottom row is the qubits $\bm{r}$ corresponding to the variables $\bm{y}$, and the others are the qubits $\bm{q}$ corresponding to the variables $\bm{x}$.
\label{fig:fig5}}

In the facility location problem, $x_{i,j}$ is a binary variable that takes $1$ if the $j$th customer is assigned to the $i$th facility and $0$ otherwise. 
$y_i$ is a variable that takes $1$ if the $i$th facility is opened and $0$ otherwise.
Let $m$ and $n$ be the total number of customers and that of facilities, respectively.
As the typical constraints, a group of conditions 
\begin{equation}
\label{eq:eq11}
\sum_{i=1}^n x_{i,j} =1 \ \ \text{for} \ j=1,2,\ldots,m,
\end{equation}
\begin{equation}
\label{eq:eq12}
x_{i,j} \leq y_i  \ \ \text{for} \ i=1,2,\ldots,n, \,\, j=1,2,\ldots,m
\end{equation}
is imposed.
We let $\bm{q}$ and $\bm{r}$ be the qubits corresponding to the variables $\bm{x}$ and those to the variables $\bm{y}$, respectively.

Construction of VQCs for the facility location problem is as follows.
The auxiliary qubits $a_j \ \text{for} \ j=1,2,\ldots,m$ are utilized for forwarding operations.
We consider the case where $m\to k$ in the constraints~\eqref{eq:eq11} and \eqref{eq:eq12}.
$k$ denotes a natural number less than or equal to $m$.
A feasible intermediate ansatz for such the $k$th subproblem is represented by $\left| \psi_k\right\rangle$.
Thus, any states outputted from $\left| \psi_k\right\rangle$ must satisfy ($k-1$)th subproblem's constraints $\sum_{i=1}^{n} q_{i,j} =1 \ \ \text{for} \ j=1,2,\ldots,k$ and $q_{i,j} \leq r_i \ \ \text{for} \ i=1,2,\ldots,n$ and $j=1,2,\ldots,k$.
In the interpolative case of $k=0$, that is the case where the variables $\bm{x}$ are absent, the constraints are trivial. The resolution of constraints ensures that $y_1,y_2,\ldots,y_{n}$ have arbitrary values.
Thus, $\left| \psi_0\right\rangle$ can be prepared by acting $R_y (\phi_i)$ gates on $r_{1},r_{2},\ldots,r_{n}$. Here, $\phi_i$ is a variational parameter.
Next, the forwarding operation that generates $\left| \psi_k\right\rangle$ from $\left| \psi_{k-1}\right\rangle$ is explained. 
The qubits are rearranged by the following matrix form 
\begin{equation}
\label{eq:eq13}
\left(\begin{matrix}
q_{1,k}&q_{2,k}&\cdots&q_{n,k}&a_{k}\\
q_{1,k-1}&q_{2,k-1}&\cdots&q_{n,k-1}&a_{k-1}\\
\vdots&\vdots&\ddots&\vdots&\vdots\\
q_{1,2}&q_{2,2}&\cdots&q_{n,2}&a_{2}\\
q_{1,1}&q_{2,1}&\cdots&q_{n,1}&a_{1}\\
r_{1}&r_{2}&\cdots&r_{n}&*
\end{matrix}\right).
\end{equation}
Here, we introduce the $*$ element for convenience, and no qubits are allocated in this element.
In the left of Fig.~\ref{fig:fig5}, $q_{1,1},\ldots,q_{n,k-1},r_1,\ldots,r_{n}$ satisfying ($k-1$)th subproblem constraints are given inside the black frame.
An one-hot bitstring ($q_{1,k},q_{2,k},\ldots,q_{n,k}$) and all-ones bitstring ($a_{k}$) are added, as shown in the red and green frame, respectively.
In Fig.~\ref{fig:fig5}, the specific case where $q_{1,k} = 1$ is depicted.
The second constraint is violated because $r_1$ remains $\left| 0\right\rangle$.
By swapping the bitstring inside the blue frame and that inside the green frame, this constraint is recovered with the other constraints preserved.
Thus, $q_{1,1},\ldots,q_{n,k},r_1,\ldots,r_{n}$ can satisfy $k$th subproblem's constraints.

The explicit gates are as follows. 
First, $\left| \psi_0\right\rangle$ is obtained  by acting $R_y (\phi_i)$ gates on $r_{1},r_{2},\ldots,r_{n}$. 
Then, forwarding operations are implemented. A parameterized W state is prepared on $q_{1,k},q_{2,k},\ldots,q_{n,k}$. $a_{k}$ is initialized to $\left| 1\right\rangle$ with an $X$ gate.
After $q_{u,k} \ \text{for}\ u=1,2,\ldots,n$ are set as control bits, CSWAP gates act on $r_{u}$ and $a_{k}$.
These gates can generate $\left| \psi_k\right\rangle$.
Performing this forwarding operation up to $k=m$ creates the target state $\left| \psi_ m\right\rangle$, which represents a feasible ansatz for the original problem.

We prove that $\left| \psi_ m\right\rangle$ is fully feasible as follows. 
By preparing parameterized W states at every rows, all possible bitstrings of the variables $\bm{x}$ are encoded.
In addition, proposed method enumerates $\bm{y}$ satisfying the second constraint for each possible bitstring of $\bm{x}$. Thus, the ansatz $\left| \psi_ m\right\rangle$ can generate states superpositioned by all feasible solutions.

In the proposed method, the number of qubits is $mn+n+m$.
This number can be reduced into $mn+n+1$ by utilizing the reuse of qubits~\cite{michael2021init}.
However, in the later experiment, this technique is not adopted because the simulation time may increase with the initialization process of qubits.

\subsection{Remarks on Other Constraints}
\label{subsec:other}
The proposed method can be applied to other constraints.
For example, the proposed method can handle one-dimensional type constraints such as $\Pi_{i=1}^n x_i =y$, which is used for dimensionality reduction~\cite{salvador2018qa}.
Let $\left| \psi_k \right \rangle$ be a fully feasible ansatz for subproblems with $k=1,2,\ldots,n$ and $r$, $q_i$, and $a$ be the qubits corresponding to $y$, $x_i$, and an auxiliary bit, respectively.
The $k$th forwarding operation $U_k$ from $\left| \psi_{k-1}\right\rangle$ to $\left| \psi_{k}\right\rangle$ is as follows.
First, $a$ is initialized to $\left| 0 \right\rangle$.
When $q_{k-1}$ in $\left| \psi_{k-1}\right\rangle$ is $\left| 0 \right\rangle$, $r$ and $a$ are swapped so as to ensure $k$th subproblem's constraint. This operation can be implemented with $X$ gates on $q_{k-1}$ and a CSWAP gate on $r$ and $a$. 

The proposed method has the capacity to address both cardinality constraints and one-hot constraints.
To illustrate this capacity, consider the modified facility location problem in Subsection~\ref{subsec:feasibleVQC_facility}, where the constraint~\eqref{eq:eq11} is generalized to $\sum_{i=1}^n x_{i,j}=d$ where $d$ is a natural number.
It is widely recognized that QAOA ansatzes with an $XY$ mixer~\cite{wang2020qaoa} can handle such a cardinality constraint.
Thus, forwarding operation utilizing $XY$ mixers instead of parameterized W states can derive the fully feasible VQC.

\section{Circuit Costs}
\label{sec:cost}

\begin{table*}[h]
    \centering
    \caption{Summary of VQC costs: the number of qubits, the number of variational parameters, and quantum gate costs. Here, quantum gate costs are the number of two-bit CNOT gate.
    These costs do not consider additional gates for measuring expected energies. When the expected values are calculated by classical computers, the additional gates are not required. }
    \label{table:table1}
    \begin{tabular}{cccc}
        \hline \hline
        Problem & Resource ($\sharp$) & $l$-layer VQE & Proposed\rule[0pt]{0pt}{10pt} \\ \hline
         & Qubits & $mn$ & $mn$ \rule[0pt]{0pt}{10pt}\\ \cline{2-4}
        Assignment ($n$ persons, $m$ jobs, $m \le n$) & Variational parameters & $(l+1)mn$ & $mn-m^2/2-m/2$ \rule[0pt]{0pt}{10pt}\\ \cline{2-4}
         & CNOT gates & $\mathcal{O}(lmn)$ & $\mathcal{O}(7m^2n/2-7m^3/6)$ \rule[0pt]{0pt}{10pt}\\
        \hline
         & Qubits & $mn+n$ & $mn+n$ \rule[0pt]{0pt}{10pt}\\ \cline{2-4}
        Shift scheduling ($n$ persons, $m$ jobs, $m \le n$) & Variational parameters & $(l+1)(mn+n)$ & $mn-m^2/2+n-3m/2$ \rule[0pt]{0pt}{10pt}\\ \cline{2-4}
         & CNOT gates & $\mathcal{O}(lmn)$ & $\mathcal{O}(7m^2n/2-7m^3/6)$ \rule[0pt]{0pt}{10pt}\\
         \hline
         & Qubits & $mn+m$ & $mn+n+m$ \rule[0pt]{0pt}{10pt}\\ \cline{2-4}
        Facility location ($n$ persons, $m$ facilities)& Variational parameters & $(l+1)(mn+m)$ & $mn+n-m$ \rule[0pt]{0pt}{10pt}\\ \cline{2-4}
         & CNOT gates & $\mathcal{O}(lmn)$ & $\mathcal{O}(9mn)$ \rule[0pt]{0pt}{10pt}\\
         \hline
    \end{tabular}
\end{table*}

In the previous section, fully feasible VQCs were constructed for three typical constrained combinatorial optimization problems. 
For each circuit, the number of required qubits, that of variational parameters, and the cost of the quantum circuit are estimated, and the results are listed in Table~\ref{table:table1}. 
The results for conventional $l$-layer VQE~\cite{amaro2022jss}, in which a layer of CNOT gates and that of $R_y$ gates are repeated $l$ times after $R_y$ gates are initially applied, are also enumerated.
Here, conventional VQEs use the penalty function method. In other words, the expected values of the cost Hamiltonian with penalty terms for the constraints are measured so as to update the variational angle of the rotation gate.
The penalty terms for each problem are as follows.
Those for the constraints~\eqref{eq:eq5}, \eqref{eq:eq8}, and \eqref{eq:eq11} are determined by the squared term $\lambda(\sum_{i=1}^n x_{i,j}-1)^2$. 
According to the literature~\cite{quinones2020pqc}, $\lambda(\sum_{j=1}^m \sum_{k<j}^m x_{i,j} x_{i,k})$ and $\lambda x_{i,j }(1-y_i)$ are adopted for the constraints~\eqref{eq:eq6} and \eqref{eq:eq12}, respectively. 
In addition, because the constraint~\eqref{eq:eq9} is equivalent to the AND condition of \eqref{eq:eq6} and \eqref{eq:eq12}, the penalty terms for \eqref{eq:eq9} can be obtained by summing the above terms. 
Thus, for the three typical problems presented in the previous section, the penalty terms can be formulated without slack variables. In other words, no auxiliary qubits are required.

As shown in Table~\ref{table:table1}, the number of required qubits is the same as that of the conventional $l$-layer VQE, or slightly larger. Specifically, our method utilizes auxiliary qubits in the facility location problem, thereby increasing the number of qubits. 
However, when the auxiliary qubit is initialized at each time of the forwarding operation, the increase can be avoided, as described in Subsection~\ref{subsec:feasibleVQC_facility}.
On the other hand, the numbers of variational parameters are reduced in all cases, which is expected to enable efficient learning of a VQC. The detail of calculating the numbers is referred to Appendix~\ref{sec:num_para}.

Next, the cost of gates in quantum circuits is discussed. In this study, the number of CNOT gates is considered. 
Because a parameterized W state and CSWAP gates are frequently used in the VQCs proposed in Section~\ref{sec:inductive}, we first calculate the cost of these gates. 
A $d$-qubit parameterized W state can be decomposed into one $X$ gate and $2d-2$ each of $H$ gates, rotation gates, and CNOT gates.
In addition, a CSWAP gate can be decomposed into 
nine $\mathrm{SU}(2)$ gates and seven CNOT gates~\cite{cruz2024cswap}.

The gate costs are calculated for the assignment problem. The $k$th forwarding operation $U_k\ (k=2,\ldots,m)$ adds a parameterized W state with the dimension of $d=n-m+k$. In addition, preparation of $\left| \psi_{1}\right\rangle$ requires that state with the dimension of $d=n-m+1$.
Thus, the number of CNOT gates required to generate all parameterized W states is at most $\sum_{k=1}^m (2d-2)=2mn-m^2-m$.
Because a CSWAP gate is used $(n-m+k-1)(k-1)$ times in the $k$th forwarding operation $U_k$, the number of CNOT gates is at most $\sum_{k=2}^m (n-m+k-1)(k-1)\times 7=7m^2 n/2-7m^3/6-7mn/2+7m/6$. 
By summing the above results, as the gate costs for the assignment problem, we obtain the total number of CNOT gates: at most $7m^2 n/2-7m^3/6-3mn/2-m^2+m/6$.
Similarly, the costs are calculated for the other problems in Appendix~\ref{sec:num_cnot}. Only the leading order of them is listed in Table~\ref{table:table1}. Note that, in the facility location problem, whereas the problem with $m$ customers and $n$ facilities was considered for convenience of explanation in Subsection~\ref{subsec:feasibleVQC_facility}, we use $n$ customers and $m$ facilities for the sake of comparison with other problems in this table.

Compared with the conventional $l$-layer VQE, the assignment and shift scheduling problems require the polynomial gate costs of $3/2$ power of the number of qubits $mn$. In the facility location problem, the costs are linear with the number of qubits, and comparable to those of the conventional VQE when the number of layers $l$ is treated as a constant.
Especially, for CNOT gates which are generally considered expensive, the cost will be lower than that for deep VQEs. In other words, a VQC customized to output only feasible solutions can be more hardware-efficient than a default VQE without such a filtering mechanism. This is because the number of CSWAP gates per forwarding operation is moderate.

\section{Experiment}
\label{sec:exp}
\Figure[t!](topskip=0pt, botskip=0pt, midskip=0pt)[width=10.5cm]{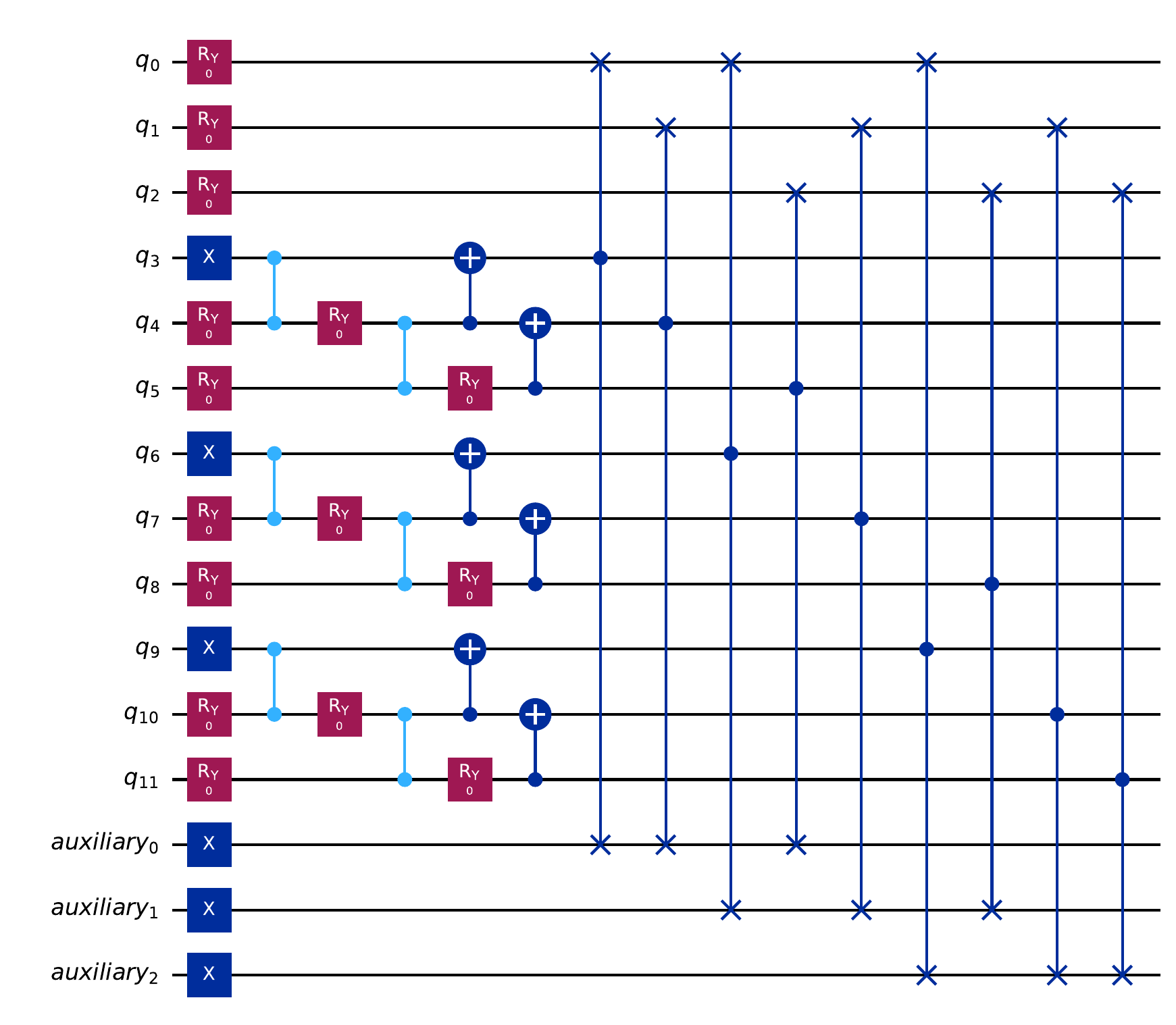}
{Fully feasible VQC for facility location problem with $m=n=3$. The circuit is constructed according to Subsection~\ref{subsec:feasibleVQC_facility}.
$q_0,q_1,q_2$ and $q_3,q_4,\ldots,q_{11}$ denote the qubits corresponding to $y_1,y_2,y_3$ and $x_{1,1},x_{2,1},\ldots,x_{3,3}$, respectively. 
Three parameterized W states are prepared on $q_3,q_4,\ldots,q_{11}$ and three auxiliary qubits are used for CSWAP gates in forwarding operations. 
Whereas there are $15$ rotation gates, the number of independent variational parameters is $9$ because the angles of the two rotation gates are zero-sum in each register of $q_4, q_5, q_7, q_8, q_{10}, q_{11}$, as detailed in Appendix~\ref{sec:pws}.
Here, for simplicity, all the values of the variational parameters are set to zeros.
The picture was generated by Qiskit's draw method~\cite{2023qiskit}.
\label{fig:fig6}}

For numerical experiments, the proposed method is applied to the facility location problem. 
Whereas a quantum-inspired approach using tensor networks~\cite{nakada2024tn} and QAOA based approach~\cite{wang2024qaoa} have been reported as related studies on this problem, the proposed method is the first time to derive fully feasible VQCs. 
Let $x_{i,j}$ be a binary variable that takes $1$ if the $j$th customer is assigned to the $i$th facility and $0$ otherwise,
and $y_i$ be a variable that takes $1$ if $i$th facility is opened and $0$ otherwise. In this study, 
\begin{equation}
\label{eq:eq14}
H(\bm{x},\bm{y})=\sum_{i=1}^n A_i y_i + \sum_{i=1}^n \sum_{j=1}^m B_{i,j} x_{i,j}
\end{equation}
is assumed as cost Hamiltonian corresponding to an objective function. 
$m$ and $n$ denote the number of customers and that of facilities, respectively. The constraints are defined by \eqref{eq:eq11} and \eqref{eq:eq12}. In the following, we consider the case where $m=n=3$.

The simulation setup is described. As the circuit simulator, Qiskit's statevector simulator~\cite{2023qiskit} was used.
The VQC was constructed as described in Subsection~\ref{subsec:feasibleVQC_facility} and its schematic picture is shown in Fig.~\ref{fig:fig6}. The variational parameters were optimized to minimize the expected energy of the cost Hamiltonian~\eqref{eq:eq14}. The expected values were calculated by averaging the energies over $2000$ shots. COBYLA~\cite{powell1994cobyla} was adopted to search for the variational parameters and SciPy library~\cite{2020scipy} was used for its calculation. 
The initial values of the search were generated from $[0,2\pi]$ with a uniform distribution and the maximum number of iterations was set to $300$.
\begin{table*}[t!]
    \caption{Results of optimization for facility location problem. Probability  ($\%$) of measuring feasible solutions and that of measuring optimal solutions are calculated from sampled states. As conventional methods, $l$-layer VQEs ($l=1,2,3$) were adopted. In these baselines, penalty function method was used and its penalty coefficient was set to $\lambda=5,10,15,20$. }
    \label{table:table2}
        \centering
        \begin{tabular}{cccccccccccccccc}\hline \hline
        \multicolumn{1}{c}{Method} & \multicolumn{4}{c}{1-layer VQE} & & \multicolumn{4}{c}{2-layer VQE} & & \multicolumn{4}{c}{3-layer VQE} & Proposed \rule[0pt]{0pt}{10pt}\\
        \cline{2-5}
        \cline{7-10}
        \cline{12-15}
        \multicolumn{1}{c}{$\lambda$} &
             $5$ & $10$ & $15$ & $20$ & & $5$ & $10$ & $15$ & $20$ & & $5$ & $10$ & $15$ & $20$ &  -  \rule[0pt]{0pt}{10pt}\\ \hline
            Feasible sols. &58.73 &80.65 &82.80 &81.67 & &53.40 &69.38 &71.52 &71.96 & &34.74 &48.30 &50.70 &48.88 &\textbf{100.00} \rule[0pt]{0pt}{10pt}\\
            Optimal sols.  &2.84 & 1.16 &0.93 &0.24 & &0.36 &0.72 &0.69 &0.21 & &0.64 &0.21 &0.14 &0.33 &\textbf{62.91} \rule[0pt]{0pt}{10pt}\\ \hline
        \end{tabular}
\end{table*}

\Figure[t!](topskip=0pt, botskip=0pt, midskip=0pt)[width=16.5cm]{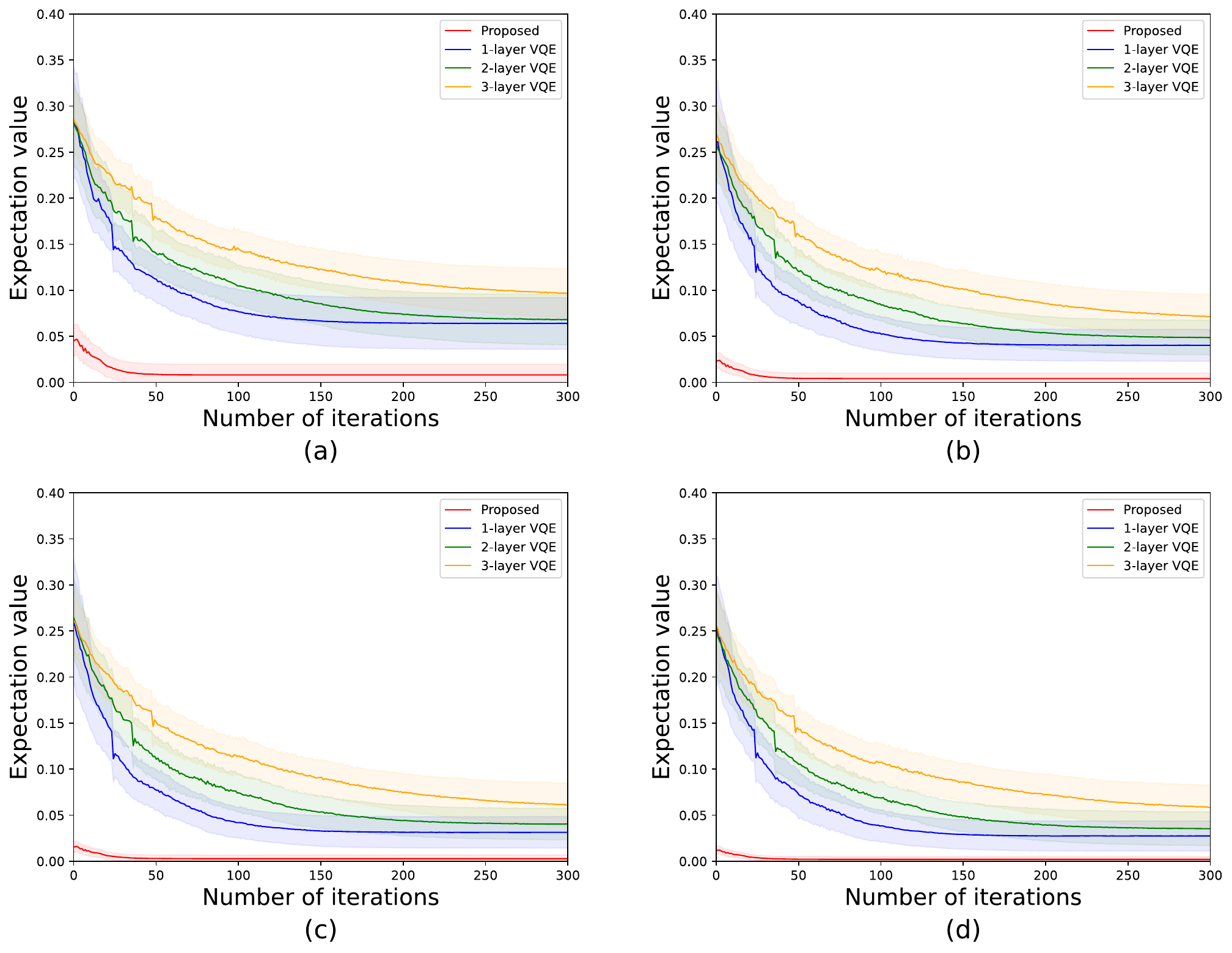}
{Changes in the values of expected energies during the VQE iteration under (a) $\lambda=5$, (b) $\lambda=10$, (c) $\lambda=15$, and (d) $\lambda=20$. The averages of the expected values in all instances are plotted as solid lines. These lines shall guide the eye. Their standard deviations are depicted in colored areas around the lines. The expected values were normalized by \eqref{eq:eq15}. Note that the results of proposed method appear to be different among multiple penalty coefficients because of the normalization.
\label{fig:fig7}}

The numerical experiments were also performed using conventional VQEs. The penalty terms for \eqref{eq:eq11} and \eqref{eq:eq12} were added to \eqref{eq:eq14} as the penalized cost Hamiltonian, and the penalty coefficient was set to $\lambda=5,10,15,20$. This is because we experimentally found that the ground states of the penalized Hamiltonian were occasionally shifted from the optimal solutions of \eqref{eq:eq14} around $\lambda=4$.
In these penalty coefficients $\lambda=5,10,15,20$, we confirmed that the optimal solutions were obtained by brute force.
$l$-layer models ($l=1,2,3$) were used as VQCs, and the other simulation settings were the same as those of the proposed method.
$A_i$ and $B_{i,j}$ were randomly sampled from $\{1,2,3,4,5\}$ and $\{0,1,2\}$ to generate $100$ instances, and optimization was performed using the proposed method and the conventional methods. The changes of the expected energies are shown in Fig.~\ref{fig:fig7}.
Here, the expectation of the cost Hamiltonian $\braket{H}$ was calculated by the states outputted from the VQC, and then was min-max normalized as follows
\begin{equation}
\label{eq:eq15}
\epsilon_{\psi} = \frac{\braket{H}-E_{\mathrm{min}}}{E_{\mathrm{max}}-E_{\mathrm{min}}},
\end{equation}
where $E_{\mathrm{min}}$ and $E_{\mathrm{max}}$ are the minimum and maximum values of the penalized Hamiltonian. These values were calculated by brute force.

From Fig.~\ref{fig:fig7}, the initial expected energies were significantly different between the proposed method and the conventional methods.
This is because the proposed method drastically reduced the value by designing VQCs to output only feasible solutions.
It was also found that the expected value converged with a smaller number of iterations than the conventional methods.
For conventional methods, the performance generally tended to deteriorate as the number of layers $l$ increases. 
This phenomenon was also observed in the paper~\cite{matsuo2023pqc}.
This may be due to the fact that an increase in the number of variational parameters raises the training difficulty.
However, the proposed method can reduce the number of parameters, which may contribute to efficient learning.

The quality of the solutions was then evaluated using the circuit obtained after the maximum iterations.
Specifically, the probability of measuring a feasible solution and that of measuring an optimal solution were estimated. To obtain these probabilities, we checked the rates of the number of such solutions in $2000$ shot measurements, then averaged them over all instances. 
The results are shown in Table~\ref{table:table2}. 
Although the number of feasible solutions was found to increase roughly by increasing the penalty coefficient $\lambda$ in conventional methods, infeasible solutions were inevitable.
This is because whereas any states outputted from the proposed VQC are feasible, the penalty term does not ensure to completely eliminate infeasible solutions.
In conventional methods, although the probability of measuring a feasible solution occasionally exceeded $80\%$, that of obtaining an optimal solution was reduced to less than a few $\%$. 
On the other hand, the proposed method succeeded in obtaining an optimal solution with more than $60\%$. 
This is considered to be due to the fact that the search space is significantly reduced by searching only for the states that satisfy the constraints. This reduction becomes more prominent as the problem size increases.

In Table~\ref{table:table2}, the strong dependency on penalty coefficient $\lambda$ was found. While increasing $\lambda$ roughly improved the measurement probability for the feasible solutions, that for the optimal solutions sometimes decreased. Furthermore, as mentioned above, the excessive small $\lambda$ has the risk of the discrepancy between the optimal solution in the original problem and the ground state in the penalized Hamiltonian. Thus, adjusting $\lambda$ is generally challenging. The proposed method has the advantage of not requiring such an adjustment.

As a result, the proposed method was found to have the highest probability of measuring the feasible and optimal solutions without any penalty function methods. In this experiment, only one initial set of variational parameters was given for each instance. We note that the probability of measuring optimal solutions can be further improved by increasing the number of initial sets.

\section{Summary and Discussion}
\label{sec:summary}
In this paper, we propose improved VQCs for a wider variety of constrained combinatorial optimization problems. 
The proposed method utilizes a forwarding operation that maps from feasible states for subproblems to those for larger subproblems. 
After a VQC for the smallest subproblem is constructed, a forwarding operation is added sequentially.
This process allows inductive construction of fully feasible VQCs even for multiple and complex constraints, as long as the appropriate forwarding operation is designed. 
As examples of combinatorial optimization problems with typical constraints, we construct VQCs for assignment problem, shift scheduling problem, and facility location problem. 
Our study is the first study to derive fully feasible VQCs for these problems.
Specifically, in the facility location problem, the cost of obtained circuits is comparable with that of a conventional VQE without any mechanisms of feasible solution filtering. 
Numerical simulations were performed for the facility location problem to verify the effectiveness of the proposed method. 
As a result, feasible and optimal solutions were found with a higher probability than conventional VQE methods.

Future issues are discussed. One drawback of the method is the difficulty in applying it to the constraint structures for which forwarding operations cannot be defined. Thus, extension to more general constraint problems is desired.
In addition, warm-starting QAOA has been proposed recently for better solutions~\cite{egger2021wsqaoa}. 
Our fully feasible VQCs can work as the initial states for such solvers. Thus, improving QAOA with the proposed method would be an interesting future work.

\appendices
\section{Parameterized W State}
\label{sec:pws}
\Figure[h!](topskip=0pt, botskip=0pt, midskip=0pt)[width=12cm]{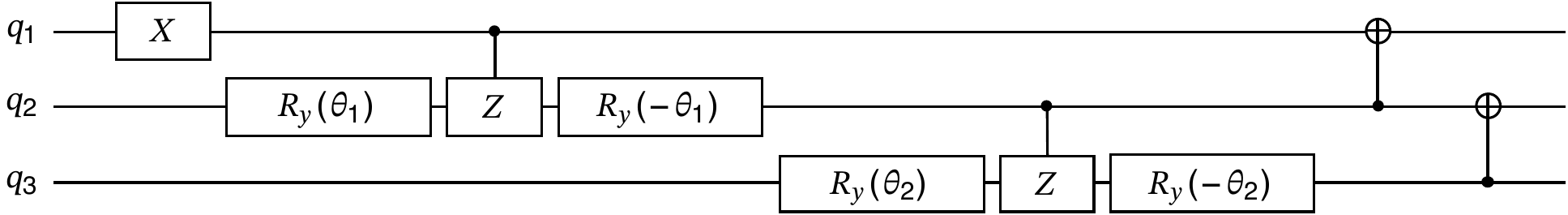}
{VQC for parameterized W state. The circuit generates mixed states of $\left| 100 \right\rangle, \left| 010 \right\rangle$, and $\left| 001 \right\rangle$.
The coefficients are not equalized but parameterized by the angles of rotation gates.
\label{fig:fig8}}
W states are equally superpositioned by one-hot states~\cite{diker2022wstate}.
However, in VQE algorithms, each amplitude must be controlled by variational parameters.
Thus, a parameterized W state was proposed~\cite{matsuo2023pqc}. Fig.~\ref{fig:fig8} shows the VQC for the case of three qubits. The output state of this circuit is $\cos{\theta_1}\left| 100 \right\rangle-\sin{\theta_1}\cos{\theta_2}\left| 010 \right\rangle+\sin{\theta_1}\sin{\theta _2}\left| 001 \right\rangle$.
Generally, the parameterized W state for $d$-qubit system has $d-1$ variational parameters. In this paper, this state is referred to as the parameterized W state with the dimension of $d$.

\section{Number of Variational Parameters}
\label{sec:num_para}
The number of variational parameters is calculated for the assignment problem in Subsection~\ref{subsec:feasibleVQC_assignment}. 
The initial ansatz $\left| \psi_1\right\rangle$ is obtained by preparing a parameterized W state with the dimension $d=n-m+1$. In addition, the $k$th forwarding operation $U_k\ (k=2,\ldots,m)$ adds a parameterized W state with the dimension of $d=n-m+k$. Thus, according to Appendix~\ref{sec:pws}, the number of variational parameters is $\sum_{k=1}^m (d-1)=mn-m^2/2-m/2$.

Next, the number of variational parameters is calculated for the shift scheduling problem in Subsection~\ref{subsec:feasibleVQC_shift}. 
The initial ansatz $\left| \psi_0\right\rangle$ is obtained by acting $R_y (\phi_i)$ gates on $r_{1},r_{2},\ldots,r_{n-m}$.
In addition, the $k$th forwarding operation $U_k\ (k=1,\ldots,m)$ adds a parameterized W state with the dimension of $d=n-m+k$. Thus, the number of variational parameters is $(n-m)+\sum_{k=1}^m (d-1)=mn-m^2/2+n-3m/2$.

Finally, the number of variational parameters is calculated for the facility location problem in Subsection~\ref{subsec:feasibleVQC_facility}. 
The initial ansatz $\left| \psi_0\right\rangle$ is obtained  by acting $R_y (\phi_i)$ gates on $r_{1},r_{2},\ldots,r_{n}$.
In addition, the $k$th forwarding operation $U_k\ (k=1,\ldots,m)$ adds a parameterized W state with the dimension of $d=n$. Thus, the number of variational parameters is $n+\sum_{k=1}^m (d-1)=mn+n-m$.

\section{Number of CNOT Gates for Shift Scheduling Problem and Facility Location Problem}
\label{sec:num_cnot}
The gate costs are calculated for the shift scheduling problem in Subsection~\ref{subsec:feasibleVQC_shift}. The $k$th forwarding operation $U_k\ (k=1,\ldots,m)$ adds a parameterized W state with the dimension of $d=n-m+k$. 
Thus, the number of CNOT gates required to generate all parameterized W states is at most $\sum_{k=1}^m (2d-2)=2mn-m^2-m$.
Because a CSWAP gate is used $(n-m+k-1)k$ times in the $k$th forwarding operation $U_k$, the number of CNOT gates is at most $\sum_{k=1}^m (n-m+k-1)k\times 7=7m^2 n/2-7m^3/6+7mn/2-7m^2/2-7m/3$. 
By summing the above results, we obtain the total number of CNOT gates: at most $7m^2 n/2-7m^3/6+11mn/2-9m^2-10m/3$.

Next, the gate costs are calculated for the facility location problem in Subsection~\ref{subsec:feasibleVQC_facility}. The $k$th forwarding operation $U_k\ (k=1,\ldots,m)$ adds a parameterized W state with the dimension of $d=n$. 
Thus, the number of CNOT gates required to generate all parameterized W states is at most $\sum_{k=1}^m (2d-2)=2mn-2m$.
Because a CSWAP gate is used $n$ times in the $k$th forwarding operation $U_k$, the number of CNOT gates is at most $\sum_{k=1}^m n\times 7=7mn$. 
By summing the above results, we obtain the total number of CNOT gates: at most $9mn-2m$.

\section*{Acknowledgment}
S. TANAKA wishes to express their gratitude to the World Premier International Research Center Initiative (WPI), MEXT, Japan, for their support of the Human Biology-Microbiome-Quantum Research Center (Bio2Q).

\vspace{-30pt}
\begin{IEEEbiography}[{\includegraphics[width=1in,height=1.25in,clip,keepaspectratio]{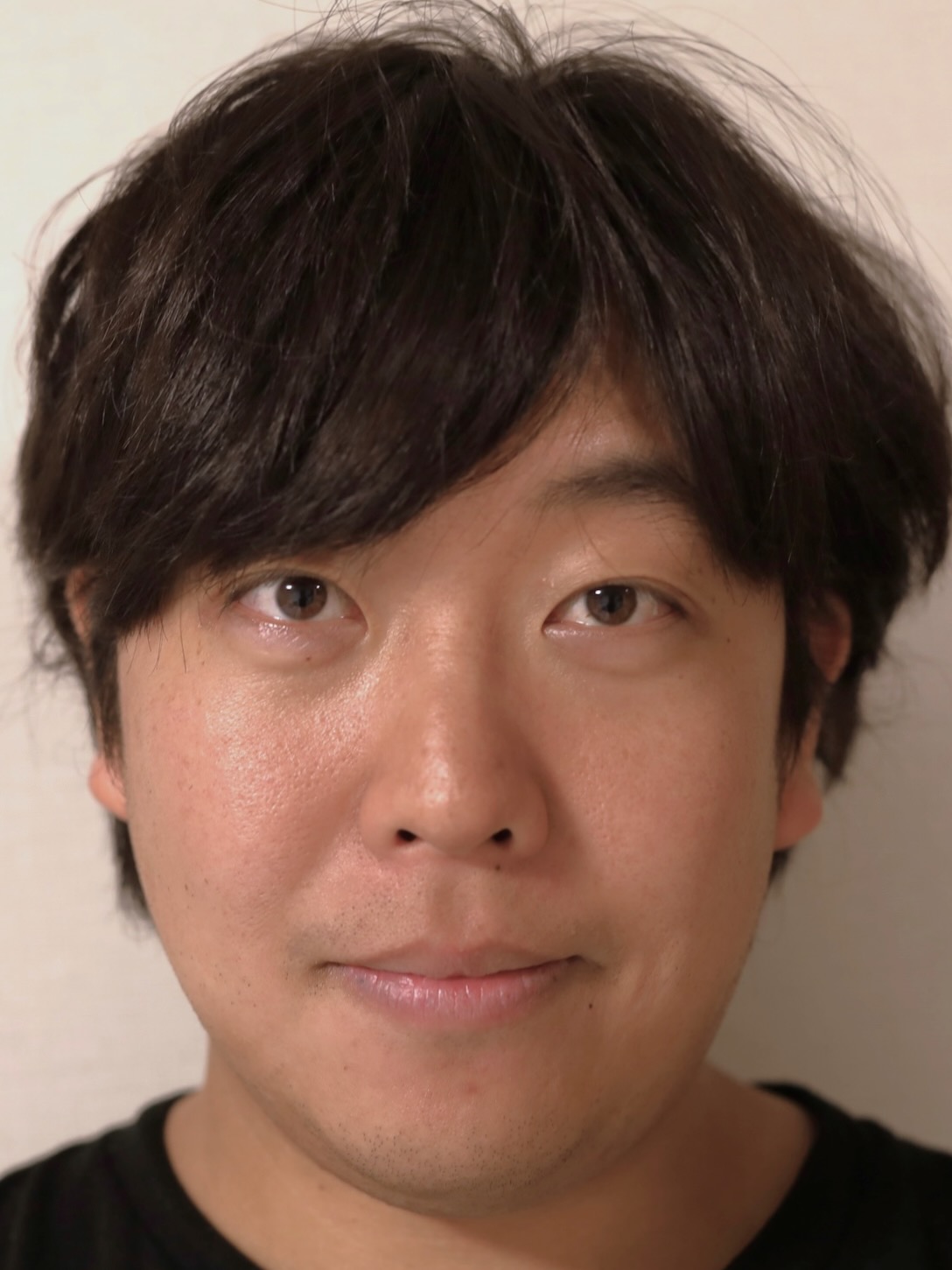}}]{Hyakka Nakada}
received his B. Sci. and M.
Sci. degrees from The University of
Tokyo in 2014 and 2016, respectively. 
He is currently pursuing a Ph.~D. degree in applied physics at Keio University.
He is also working for Recruit Co., Ltd., Tokyo, Japan. 
His research interests
include quantum computing, statistical mechanics, and machine learning.
He is a member of
the Physical Society of Japan (JPS), and The Japan Statistical Society (JSS).
\end{IEEEbiography}

\vspace{-20pt}
\begin{IEEEbiography}[{\includegraphics[width=1in,height=1.3in,clip,keepaspectratio]{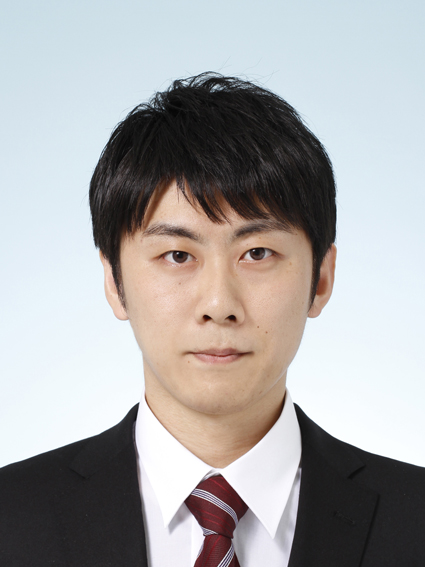}}]{Kotaro Tanahashi}
received his M.~Eng. degree from Kyoto University, Kyoto, Japan, in 2015. He has experience working as a machine learning engineer at Recruit Co., Ltd., and is working for Turing Inc., Tokyo, Japan. He also serves as a project manager for MITOU Target Program of the Information-Technology Promotion Agency (IPA).
His research interests include mathematical optimization, quantum annealing, Ising machines, machine learning, and autonomous driving.
\end{IEEEbiography}
\vspace{-20pt}
\begin{IEEEbiography}[{\includegraphics[width=1in,height=1.5in,clip,keepaspectratio]{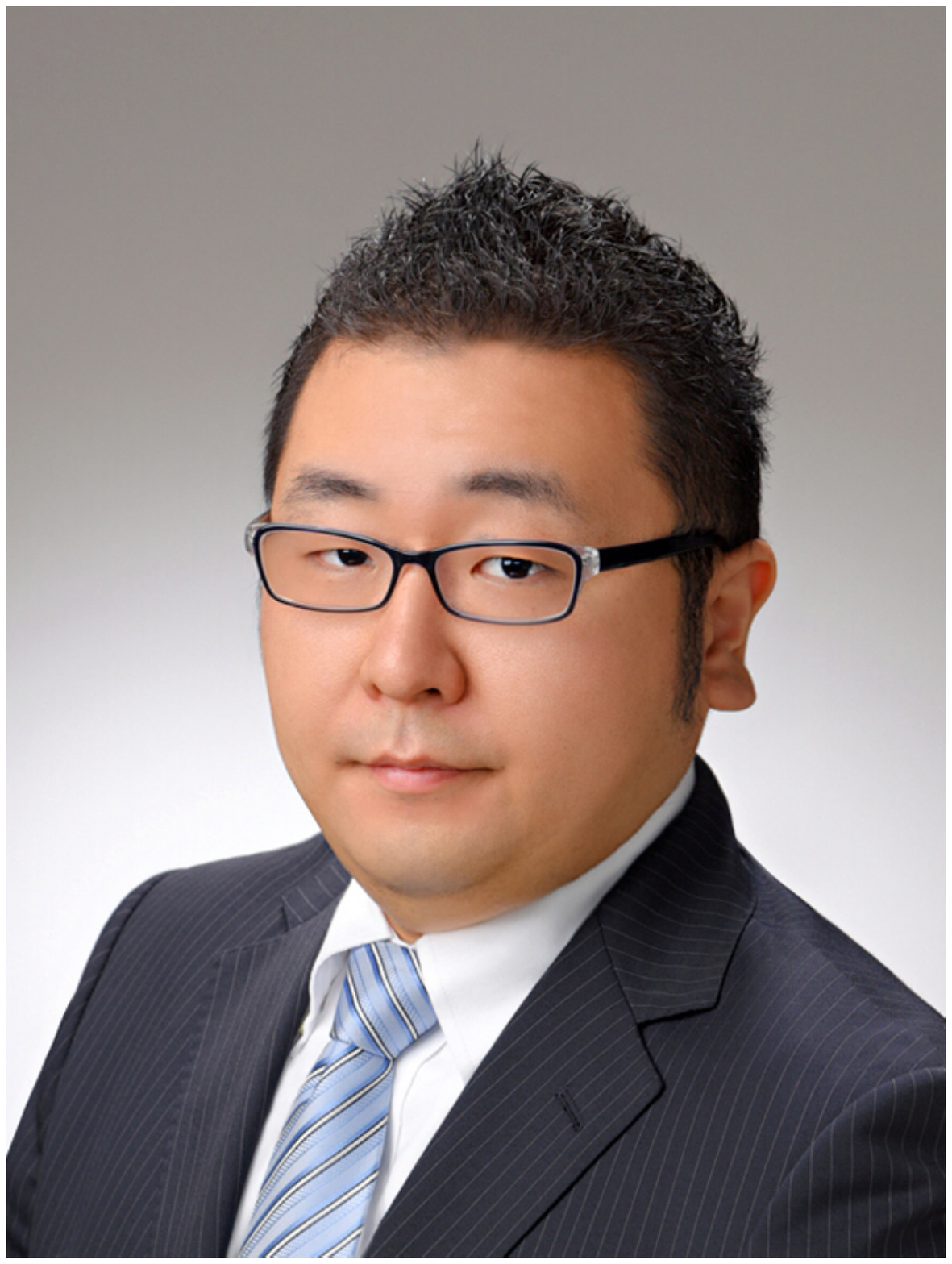}}]{Shu Tanaka} (Member, IEEE) received his B.~Sci. degree from the Tokyo Institute of Technology, Tokyo, Japan, in 2003, and his M.~Sci. and Ph.~D. degrees from the University of Tokyo, Tokyo, Japan, in 2005 and 2008, respectively.
He is currently an Associate Professor in the Department of Applied Physics and Physico-Informatics, Keio University, a chair of the Keio University Sustainable Quantum Artificial Intelligence Center (KSQAIC), Keio University, and a Core Director at the Human Biology-Microbiome-Quantum Research Center (Bio2Q), Keio University. 
He is also a visiting associate professor at the Green Computing Systems Research Organization (GCS), Waseda University. 
His research interests include quantum annealing, Ising machines, quantum computing, statistical mechanics, and materials science. 
He is a member of the Physical Society of Japan (JPS), and the Information Processing Society of Japan (IPSJ).
\end{IEEEbiography}

\EOD

\end{document}